# Heteromoiré Engineering on Magnetic Bloch Transport in Twisted Graphene Superlattices

Fanrong Lin[1,2,†], Jiabin Qiao[1,†], Junye Huang[1,3], Jiawei Liu[1,2], Deyi Fu[1], Alexander S. Mayorov[1], Hao Chen[1,2], Paromita Mukherjee[1,2], Tingyu Qu[1,2], Chorng Haur Sow[1,2], Kenji Watanabe[4], Takashi Taniguchi[4], Barbaros Özyilmaz[1,2,3*]

[1]Centre for Advanced 2D Materials, National University of Singapore, 6 Science Drive 2, 117546, Singapore

[2]Department of Physics, National University of Singapore, 2 Science Drive 3, 117551, Singapore

[3]Department of Materials Science and Engineering, National University of Singapore, 117575, Singapore

[4]National Institute for Materials Science, Namiki 1-1, Tsukuba, Ibaraki 305-0044, Japan

[†]These authors contributed equally to this work
*E-mail : barbaros@nus.edu.sg

**Abstract**

Localized electrons subject to applied magnetic fields can restart to propagate freely through the lattice in delocalized magnetic Bloch states (MBSs) when the lattice periodicity is commensurate with the magnetic length. Twisted graphene superlattices with moiré wavelength tunability enable experimental access to the unique delocalization in a controllable fashion. Here we report the observation and characterization of high-temperature Brown-Zak (BZ) oscillations which come in two types, $1/B$ and $B$ periodicity, originating from the generation of integer and fractional MBSs, in the twisted bilayer and trilayer graphene superlattices, respectively. Coexisting periodic-in-$1/B$ oscillations assigned to different moiré wavelengths, are dramatically observed in small-angle twisted bilayer graphene, which may arise from angle-disorder-induced in-plane heteromoiré superlattices. Moreover, the vertical stacking of heteromoiré supercells in double-twisted trilayer graphene results in a mega-sized superlattice. The exotic superlattice contributes to the periodic-in-$B$ oscillation and dominates the magnetic Bloch transport.

Quantum oscillations in solid-state systems have emerged as a powerful tool to unveil the novel physics of quasiparticles subject to external magnetic field ($B$)[1-5]. For example, in the presence of a uniform magnetic field, the energy spectrum of electrons in graphene lattice encounters Landau quantization where all electron states are quenched into Landau levels (LLs), manifested by electrons confined in circular orbits (Fig. 1a). As $B$ increases, the LLs are driven toward higher energy and successively pass through the Fermi level. Consequently, the periodic transition of the highest LL with respect to the Fermi level from fully occupied to empty, recorded by the decrease of LL filling factor $v$, contributes to the oscillations in magnetoresistance in $1/B$ periodicity, or the so-called Shubnikov-de Haas (SdH) oscillations[1]. Such an oscillatory behavior could be observed not only in atomic lattices but also in artificially designed materials with larger lattices, such as graphene/hexagonal boron nitride (G/hBN) moiré superlattices[4]. Due to the larger lattices, a new type of quantum oscillations independent of Landau quantization, which is referred to as Brown-Zak (BZ) oscillations[6,7], can be studied in experimentally attainable magnetic field range[4,5].

The assembly of graphene with an aligned hBN layer results in moiré superlattices with wavelength of ~ 14 nm[4,5,8-10], nearly two orders of magnitude larger than graphene lattice constant $a_0$ (~ 0.246 nm). Generally, applied magnetic fields eliminate the ability of electrons to freely propagate through the G/hBN moiré superlattices due to magnetic-field-induced localization, similar to the case in Landau-quantized graphene. However, the situation is drastically changed when the magnetic length ($l_B = \sqrt{\frac{\hbar}{eB}}$; $\hbar$, Plank's constant $h$ divided by $2\pi$) is commensurate with the moiré wavelength, corresponding to the specific values of a dimensionless ratio $\phi_0/\phi = q/p$, where $\phi = BS$ is the magnetic flux per superlattice unit cell $S$, $\phi_0 = h/e$ is the magnetic flux quantum, $p$ and $q$ are co-prime integers. Under this condition, electrons are driven into magnetic Bloch states (MBSs) and recover the access to unimpeded propagation, as if experiencing zero magnetic field (Fig. 1b)[4,5]. The MBSs, labelled by the ratio values of $\phi_0/\phi$, are responsible for periodic occurrence of minima in magnetoresistance at $B = \phi_0 p/Sq$ at high temperatures ($T$), thus leading to the BZ oscillations. Such oscillatory behavior has always been observed in (nearly) aligned G/hBN superlattices with limited moiré sizes (11 ~ 14 nm)[4,5]. In contrast, twisted graphene system, such as twisted bilayer graphene (TBG) where two graphene sheets are stacked with a twist angle (Fig. 1c)[11-22], could produce superlattices with widely tunable moiré wavelengths (1 ~ 100 nm) by tuning the twist angles[17,23], which makes it a preferable candidate for further investigation and application of the superlattice-dependent quantum oscillation. Recently, the observation of the oscillation has been reported in twisted double bilayer graphene systems[24].

In this Letter, we report the observation of high-$T$ BZ oscillations in twisted graphene superlattices. The oscillations are extremely sensitive to the moiré wavelength and characterized by two different kinds of periodicities. Periodic-in-1/$B$ oscillations are observed in TBG superlattices with lattice size in the order of 1 nm, which arises from electrons in integer MBSs. In this case, an integer number of moiré cells share one magnetic flux quantum. Intriguingly, the coexistence of the oscillations with different $\Delta(1/B)$ values is observed. This may result from the formation of angle-disorder-induced in-plane heteromoiré superlattices in the system. Moreover, another type of BZ oscillation in $B$ periodicity is observed in vertical heteromoiré supercells formed in double-twisted trilayer graphene (dTTG), where a new superlattice with moiré wavelength in the order of 100 nm is formed as a result of two constituent TBG junctions. The ultralarge lattice supports the emergence of fractional MBSs, featuring fractional filling of a moiré unit cell per flux quantum, contributing to the dominant periodic-in-$B$ oscillation in the dTTG system. Our results pave the way to study the magnetic Bloch transport in various superlattice systems *via* heteromoiré engineering.

Our experiment was firstly carried out on multiterminal Hall bar devices of the TBG system with twist angle $\theta$ of around 1° to 3°, corresponding to moiré wavelengths from several nanometers to tens of nanometers[13,25,26]. In the system, high-quality TBG encapsulated by two misaligned (larger than 15°) hBN layers forms a sandwich-like structure, which was fabricated by the developed "tear and stack" technique with an experimental angle accuracy within ±0.3° (see Methods for more details)[17,23]. The intentional misalignment between the graphene layer and the adjacent hBN layer eliminates potential influences from G/hBN superlattices[4]. The hBN/TBG/hBN Hall bar geometry was clamped by metal film and doped Si substrate where top and bottom gate voltages $V_{tg}$, $V_{bg}$ were applied, respectively.

We then focused on the TBG device B1 with twist angle set at ~ 2° during the preparation, and the corresponding optical image with device geometry is shown in Fig. 2a. Figure 2b shows a 2D color plot of the longitudinal resistance $R_{xx}^{\text{I}}$ (region I of B1) as a function of $B$ and charge density $n$ (mainly for electron doping) at $T$ ~ 20 K. Despite slight thermal smearing of Landau quantization compared to that at lower $T$ (Supplementary Note 1 and Fig. S1), a series of Landau fans are still visualized. The main fans spread from Dirac point (DP, $n$ = 0), and fully filled point (FFP) where the density $n_s = 4/S$ is needed to completely fill a superlattice band with four electrons per moiré unit cell with the area $S = \sqrt{3}\lambda^2/2$ ($\lambda$ is the wavelength)[17,19,27]. For the measurement on the region I, $n_s$ is determined to be $(1.15\pm0.01)\times10^{13}$ cm$^{-2}$, which translates to the moiré wavelength of $\lambda_1 \sim 6.34\pm0.06$ nm and the twist angle $\theta_1 \sim 2.22°$ $\pm0.02°$. Note that, surprisingly, numerous additional horizontal streaks appear in the vicinity of van Hove singularity (VHS, corresponding to $n \sim 5\times10^{12}$ cm$^{-2}$) where the abrupt charge inversion occurs[17,27], which is different from the case in G/hBN system[4,5] (see Supplementary Note 2 for more discussions). Such horizontal streak-like feature marks the existence of $n$-independent magneto-oscillations[4,5]. As we discuss below, the oscillatory behavior is ascribed to BZ oscillations originating from the formation of integer MBSs in small-angle TBG superlattices.

The physics of superlattice-related BZ oscillations could be better understood in a spectral diagram parameterized by $\phi_0/\phi = q/p$ [only curves corresponding first-order integer ($p = 1$) or fractional ($q = 1$) are shown for clarity][4,5], as shown in Fig. 1e. The diagram explicitly illustrates that MBSs in different-sized moiré superlattices emerge at different commensurate magnetic fields ($B = \phi_0 p/Sq$). In small-angle TBG superlattices, where the moiré wavelength $\lambda$ is around several to dozens of nanometers, the spectral diagram in range of $a_0/\lambda \sim 0.01$ to $0.1$ is completely filled with the integer $\phi_0/\phi$ curves ($\phi_0/\phi=1,2,3, ...$, marked by magenta curves in Fig. 1e). The integer part of the spectra is associated with the integer MBSs, which are characterized by an integer number of moiré unit cells sharing one flux quantum. The integer MBSs periodically appear in the specific superlattice, thus leading to the oscillations in $1/B$ periodicity (Fig. 1f). To highlight the oscillations in the Landau fan diagram (Fig. 2b), a close-up plot of $R_{xx}^{\mathrm{I}}$ near the VHS as a function of $\phi_0/\phi$ ($\phi$ is the magnetic flux per unit cell of TBG with $\theta_1$) is shown in Fig. 2c. The oscillations manifest themselves as a set of $n$- independent streaks with a uniform spacing $\Delta(\phi_0/\phi) = 1$, revealing the periodic-in-1/$B$ nature[4,5,8-10]. Moreover, the minima of $R_{xx}^I$ are located at the integer-valued $\phi_0/\phi$ as expected (Fig. 2d), suggesting that the oscillations are assigned to BZ oscillation governed by the recurring integer MBSs. Similar oscillations were also observed in the TBG device B2 with the (nearly) first magic angle $\theta \sim 1.11° \pm 0.02°$, corresponding to the wavelength $\lambda \sim 12.6 \pm 0.3$ nm (Supplementary Note 3). Additionally, the oscillations are still visible at high $T$ ($T > 100$ K) (Fig. S3), further confirming the formation of high-$T$ BZ oscillations in small-angle TBG superlattices.

Similar measurements were carried out on other regions (region II and III) of device B1. In spite that all the three regions are parts of the continuous TBG sheet which is generally assumed to possess a global and uniform superlattice (Fig. 2e), the Landau fan diagrams of region II and III exhibit dramatically distinct characteristics compared to that of region I. In region II, two sets, instead of one found in region I, of streak-like structures emerge between DP and FFP peaks, as shown in Fig. 3a. The streaks closer to DP peak are easily identified to be the same as the structures in region I, which are attributed to the BZ oscillations in the TBG superlattice with $\lambda_1$ and $\theta_1$. However, unexpectedly, the additional streaks closer to FFP peak become denser and show an unusual $n$-dependent distribution where the spacing between two adjacent streaks is almost linearly enlarged as $n$ increases. To unveil the odd oscillatory behavior, the close-up of the two sets of oscillations is shown in the color plot of $R_{xx}^{II}$ as a function of 1/$B$ (Fig. 3b). Similar to the oscillations near DP peak, two ends of the oscillations near FFP peak at $n_L \sim 8.0 \times 10^{12}$ cm$^{-2}$ and $n_R \sim 1.05 \times 10^{13}$ cm$^{-2}$ both exhibit the 1/$B$ periodicity. By linear fitting (Fig. S5), the period $\Delta(1/B)$ yields $\sim 0.0048$ T$^{-1}$ and $0.0052$ T$^{-1}$, which translate into the twist angles $\theta_L \sim 2.94°$ and $\theta_R \sim 2.82°$ of TBG superlattices at the left and right ends, respectively. Both the oscillations show uniform period $\Delta(\phi_0/\phi)_L = \Delta(\phi_0/\phi)_R = 1$ (Fig. 3b) as well, confirming the formation of different BZ oscillations at the two ends. Note that, surprisingly, the direct connection between the $i$-valued MBSs at the left end and the ($i$-1)-valued MBSs at the right end (where $i$ is a non-zero integer) perfectly captures the $n$-dependent feature, revealing that the unique oscillations may arise from the superposition of two adjacent integer MBSs (see Supplementary Note 4 for more discussions). Similar phenomena were also observed in region III. Compared to the case in region II, only one set of $n$-dependent oscillations near FFP peak emerge in region III while the oscillations near DP peak disappear, as shown in Fig. 3c. From the 1/$B$ fitting (Fig. S5), the period $\Delta(1/B)$ are determined to be ~ 0.0048 T$^{-1}$ and ~ 0.0049 T$^{-1}$, and thus the twist angles $\theta_L$~ 2.94° and $\theta_R$~ 2.89°, respectively. The $n$-dependent feature is also reproduced well by the connection between the $i$-valued MBSs on the left and the ($i$-1)-valued MBSs on the right (Fig. 3d), further confirming the superposition of different integer MBSs in the $n$-dependent oscillations.

To unveil the nature of unique oscillations observed in region II and III, we started with an analysis on the Landau fan diagram of region II where additional Landau levels fan out at higher $n$ ~ 1.99×10$^{13}$ cm$^{-2}$, which is beyond the gate doping limit (Fig. S1). The fan structures indeed reveal the emergence of the additional superlattice with smaller wavelength $\lambda_2$ (corresponding to a larger twist angle $\theta_2$) in the global TBG sheets. From the convergence point of the fainter fans, the twist angle is estimated to be $\theta_2$~ 2.92°±0.03° (Supplementary Note 1), which is close to the angles (2.82°~ 2.94°) extracted from the $n$-dependent oscillations near FFP peak in region II and III. Therefore, it is reasonable to attribute the two specific oscillatory behaviors to the formation of lateral heteromoiré superlattices, where superlattices with different moiré wavelengths or twist angles coexist in plane. Indeed, the real-space distribution of twist angle disorder is complicated and hard to map. To better understand the phenomena, a simple and effective perspective is evoked that the lateral heteromoirés may arise from the local twist angle disorder, mainly including line defects[28] and lattice heterostrain[29-31]. The coexistence of two distinct oscillations near DP and FFP peaks in region II might arise from the heteromoiré junction with two predominant wavelengths $\lambda_1$ and $\lambda_2$ induced by line defects in region II[28] (Fig. 3e, left). Moreover, the local angle could be mildly changed ($\Delta\theta < 0.3°$) *via* lattice heterostrain[29-31], resulting in the heteromoiré junction with slightly different twist angles (Fig. 3e, right). Such strain-induced heteromoiré superlattices may be responsible for the generation of *n*-dependent oscillations in region II and III.

To explore the MBSs and related oscillations in superlattices with much larger wavelengths (i.e., $a_0/\lambda \sim 0.001$ to $0.01$), the necessary twist angles in TBG are extremely tiny ($\theta \sim 0.1°$)[19,30,32-35], which is unfeasibly smaller than our experimental accuracy of angle control. Furthermore, superlattices in tiny-angle TBG experience structural reconstruction due to instability and the electronic band structures are also affected as a result[34]. Here we present an alternative method to prepare mega-sized moiré superlattices without reconstruction *via* vertical heteromoiré engineering in double-twisted trilayer graphene (dTTG) system[36-38], analogous to double-aligned hBN/G/hBN heterostructures[39-41]. In experiment, multiterminal Hall bar devices of the dTTG system were made using the same techniques as the TBG devices, where the twist angles in top two layers and bottom two layers were set at slightly different values ($0 < |\theta_t - \theta_b| < 0.3°$), thus resulting in a larger moiré superlattice with $\lambda_M \sim a_0/|\theta_t - \theta_b|$ (Supplementary Note 5).

Figure 4a shows the optical image with schematic geometry of the dTTG device T1 with $\theta_t$ and $\theta_b$ set at ~ 0.8° and ~ 0.7°, respectively. The 2D color plot of longitudinal resistance $R_{xx}$ in the device T1 reveals numerous resistance peaks as the voltages applied to the top and bottom gates $V_{tg}$ and $V_{bg}$ are tuned, as shown in Fig. 4b. Despite the complexity of the peaks arising from the complicated band structures for the dTTG system[34], two sets of triple peaks with different slopes are separately identified at high $V_{bg}$ (especially, high negative voltages). We attribute the peaks to the existence of DP and two secondary Dirac points (SDPs, for hole and electron doping) of two distinct tiny-angle TBGs formed in the top and bottom two layers, respectively[34]. The effective gate voltages applied on the two TBGs are different due to electrostatic screening effect induced by the graphene layer adjacent to the hBN sheet, leading to the differences in slope of two sets of peaks (Supplementary Note 6). Based on the electrostatic screening model, the twist angles are estimated to $\theta_t$ ~ 0.90°±0.02° and $\theta_b$ ~ 0.76°±0.02°, close to the target values. To illustrate the oscillations in the dTTG system, longitudinal resistance $R_{xx}$ as a function of $B$ and $V_{tg}$ at low $T$ (~ 6 K) was plotted as shown in Fig. 4c. We observed an unusual series of equally spaced vertical streaks, indicating the oscillations with periodicity in $B$ rather than $1/B$.

Indeed, the periodic-in-$B$ oscillatory behavior unveils the nature of BZ oscillations in the mega-sized moiré superlattices. As the moiré wavelength increases, the MBSs marked by the integer-valued $\phi_0/\phi$ converge toward lower $B$, while the MBSs with fractional $\phi_0/\phi$, featuring fractional filling of superlattice unit cell per flux quantum, become experimentally accessible and dominant the spectral diagram progressively (Fig. 1e). The periodic emergence of fractional MBSs contributes to the oscillations in $B$ periodicity (Fig. 1d), in stark contrast to 1/$B$ periodicity in integer MBSs. For the case of device T1, we extracted the oscillation period $\Delta B \sim 0.50\pm0.05$ T, which yields the moiré wavelength $\lambda \sim 98.2\pm4.9$ nm, equivalent to that of TBG with twist angle $\theta \sim 0.14°$ $\pm0.01°$. These values obtained from BZ oscillations agree perfectly with the values using the geometric analysis mentioned earlier (Fig. 4d, $\lambda_M \sim 100$ nm and $|\theta_t - \theta_b| \sim 0.14°$), suggesting that periodic-in-$B$ oscillations are indeed governed by fractional MBSs in mega-sized superlattices. Recently, similar oscillations have been experimentally observed in tiny-angle TBGs with large moiré sizes[35]. Despite both the oscillations are in $B$ periodicity, their origins are different. Giant Aharonov-Bohm (AB) oscillations with periodicity in $B$ emerge in tiny-angle TBG due to the formation of a network of 1D metallic states along AB/BA domain walls surrounding the insulating triangular domains[42,43], while the oscillations in dTTG stem from the 2D conductive mega-sized incommensurate relaxation superlattices where no standard AB/BA domain walls are technically formed to support AB oscillations[37,38].

Furthermore, our experiments reveal that there only appear the oscillations corresponding to the exotic ultra-large superlattice, in contrast to the coexistence of oscillations assigned to different superlattices in lateral heteromoiré superlattices (Fig. 3). This fact is also confirmed by measurements performed in the dTTG device T2 (Supplementary Note 7). It has been theoretically predicted that, in twisted trilayer graphene system, there is the remarkable hybridization between two sets of moiré supercells in nearly the equal twist angle regime[37,38]. Accordingly, an instructive perspective is provided that, the two different superlattices formed in top and bottom two graphene layers could be regarded as "stacked" on top of each other and constitute the vertical heteromoiré superlattices (Fig. 4d). In analogy with vertical van der Waals heterostructures[44,45], vertical heteromoiré hybridizations may result in the exotic "supermoiré" dominating the magnetic Bloch transport, which will guide further theoretical and experimental works for better understanding.

## Methods

### Device fabrication

Twisted bilayer graphene was fully encapsulated in two hexagonal boron nitride (hBN) layers by 'tear and stack' method. The heterostructure was assembled using a Poly (Bisphenol A Carbonate) film/Polypropylene Carbonate (PPC)/Polydimethylsiloxane (PDMS) stamp on a glass slide. The top surface of PDMS was treated by $O_2$ plasma under 20 sccm, 20 W for 1 minute using VITA-MINI ion etching machine to enhance the surface adhesion with PC/PPC film and avoid delamination during transfer process. The top hBN flake was picked up at 60 °C. Then using the van der Waals attraction force between top hBN flake and portion of the exfoliated monolayer graphene to tear it apart. The remaining graphene on substrate was controllably rotated and picked up sequentially at $T \sim 110$ °C. Then the picked up TBG was encapsulated with another hBN and the whole stack was transferred onto 285-nm-thick $SiO_2$ for further device fabrication.

The double twisted trilayer graphene systems were prepared by the same procedures but using 'tear and stack' method for twice.

A Ti/Au top gate was deposited on to the region defined by standard electron beam lithography, then using PMMA as a mask for $CHF_3/O_2$ etch to defined Hall bar structure. Cr/Au (5/50 nm) was subsequentially thermal deposited on to the device to form edge contacts.

### Electrical measurements

We measured the longitudinal resistance $R_{xx}$ and Hall resistance $R_{xy}$ using standard lock-in techniques with excitation frequencies of 13.373 Hz and current of 10 nA at 1.5 K ~ 10 K, and 100 nA at $T > 10$ K.

**Associated Content**

The Supporting Information is available free of charge at https://pubs.acs.org/doi/...

Extended data on magnetotransport measurements of small-angle and magic-angle TBG systems, as well as dTTG system, Origin of BZ oscillations emerging near VHS in TBG systems, theoretical calculation of two different twist angles and the supermoiré size in dTTG system, temperature dependence of BZ oscillations in $B$ and $1/B$ periodicities

**Author Information**

**Author Contributions**

[†]F.L. and J.Q. contributed equally to this work

**Note**

The authors declare no competing financial interests.

**Acknowledgements**

The authors acknowledge helpful discussions with Kostya Novoselov, Shaffique Adam, Vitor Pereira, Colin Woods, Thiti Taychatanapat, Alexander Hamilton, Stephan Roche. B.Ö. acknowledges support by the National Research Foundation, Prime Minister's Office, Singapore, under its Competitive Research Program (CRP Award No. NRF-CRP9-2011-3), NRF Investigatorship (NRFI Award No. NRF-NRFI2018-08) and Medium-Sized Centre Programme. K.W. and T.T. acknowledge support from the Elemental Strategy Initiative conducted by the MEXT, Japan and the CREST (JPMJCR15F3), JST.

**Figure**

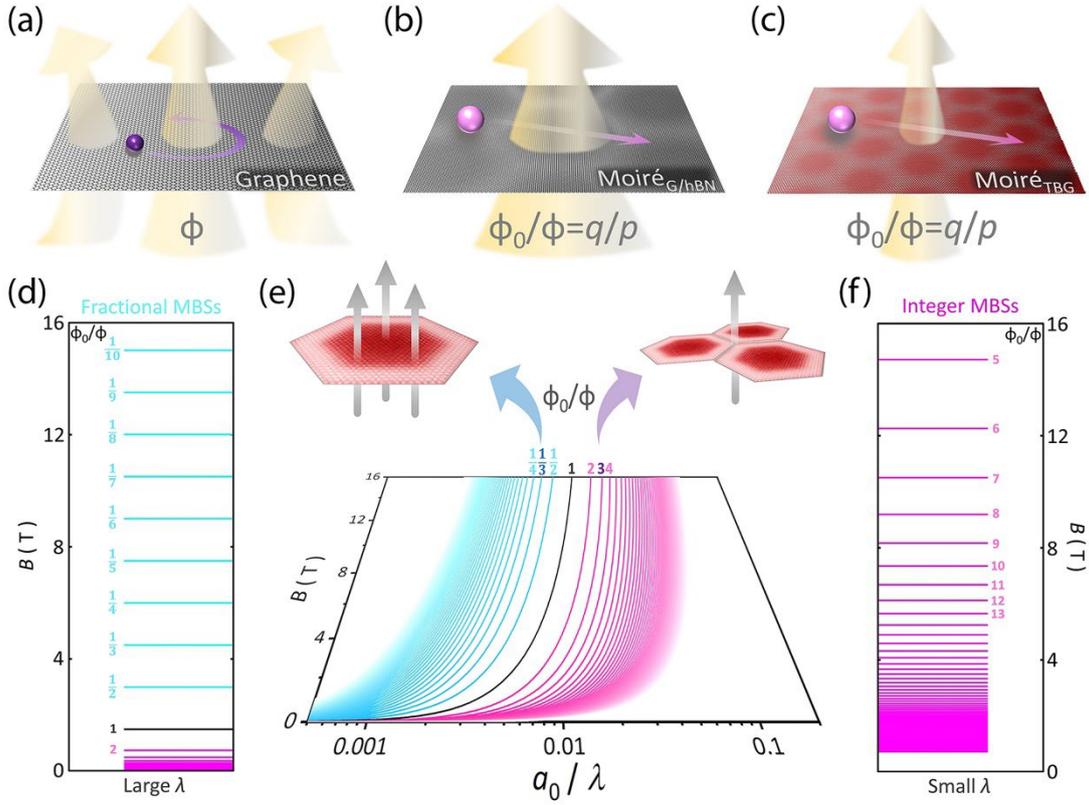

**Figure 1 | BZ oscillations originating from recurring integer and fractional MBSs in twisted graphene superlattices. (a)** Lines of magnetic flux perpendicularly passing through graphene lattice, leading to electrons (marked by the purple ball) localized in a circular orbit (marked by the purple arrow). **(b)** and **(c)** Quasiparticles (marked by the pink balls) restart to propagating along a straight trajectory (marked by the pink arrows) in G/hBN and TBG superlattices due to electrons in MBSs under the commensurate condition. **(e)** Spectral diagram of integer (the magenta curves) and fractional (the cyan curves) MBSs versus $B$ and $a_0/\lambda$. For example, the fractional and integer MBSs =1/3 (the blue sign and arrow) and 3 (the purple sign and arrow) are illustrated, which are characterized by one moiré unit cell sharing three flux quanta and three moiré cells per flux quantum, respectively. **(d)** Distribution of the fraction valued MBSs or $\phi_0/\phi$ in the superlattice with wavelength $\lambda \sim 56$ nm and twist angle $\theta \sim 0.25°$. **(f)** Distribution of

the integer valued MBSs or $\phi_0/\phi$ in the superlattice with wavelength $\lambda \sim 8$ nm and twist angle $\theta \sim 1.75°$.

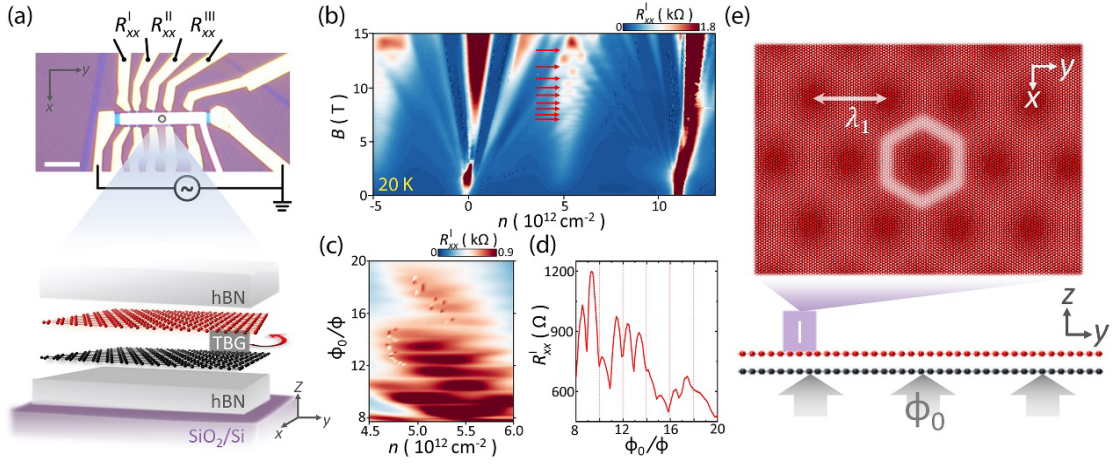

**Figure 2 | BZ oscillations governed by integer MBSs in small-angle TBG superlattices. (a)** Optical image and measurement configuration of the dual-gated TBG superlattice Hall bar device B1, where schematic of the device geometry is illustrated. Scale bar, 5 μm. **(b)** Color plot of the longitudinal resistance $R_{xx}^I$ as a function of $n$ and $B$ ($V_{bg}$ is fixed at 20 V), where the red arrows mark the dips of BZ oscillations. **(c)** Part of **(b)** near the VHS ($n \sim 5.0\times10^{12}$ cm$^{-2}$ for electron doping) is magnified and plotted as a function of $\phi_0/\phi$. **(d)** Line cut of **(c)** at $n \sim 5.25\times10^{12}$ cm$^{-2}$. **(e)** Cartoon of TBG sheet subject to external magnetic field (indicated by flux quanta) where the region I shows a uniform superlattice with the wavelength $\lambda_1$.

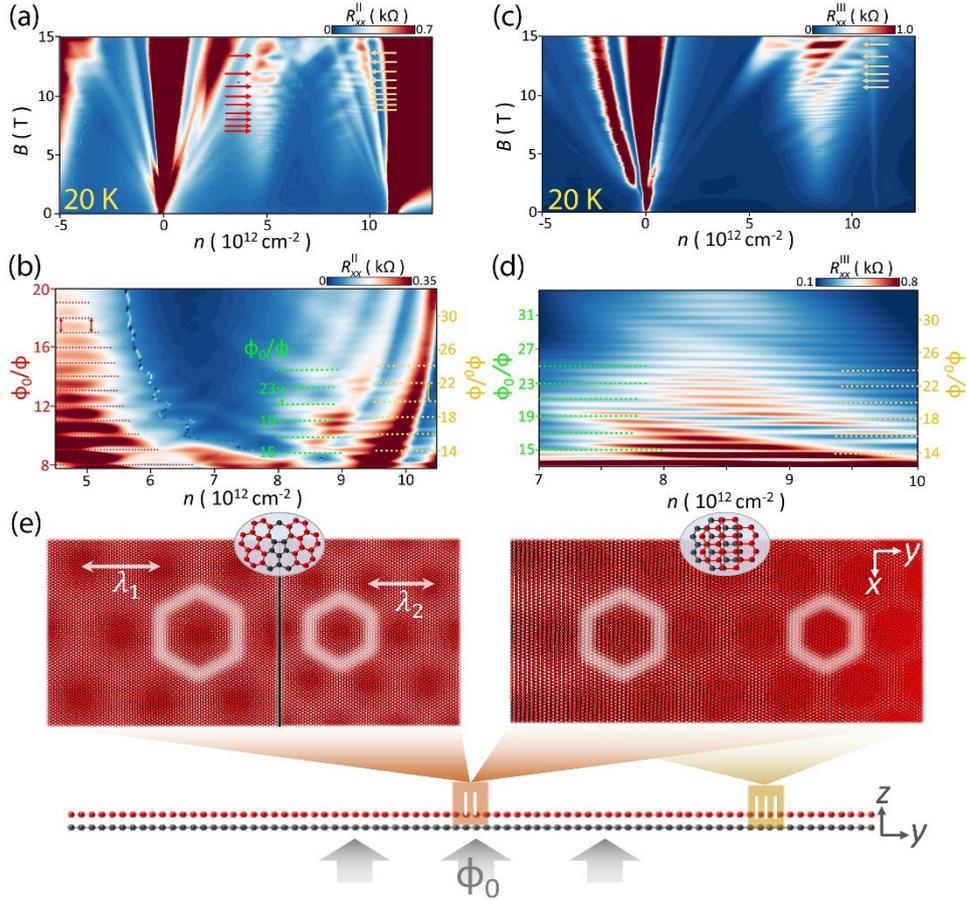

**Figure 3 | Coexistence of different BZ oscillations and superposition of two adjacent integer MBSs in lateral heteromoiré superlattices.** **(a)** Color plot of $R_{xx}^{II}$ as a function of $n$ and $B$ ($V_{bg}$ is fixed at 20 V), where the red and yellow arrows mark the $n$-independent oscillations near DP peak ($n \sim 0$) and the $n$-dependent oscillations near FFP peak ($n \sim 1.05 \times 10^{13}$ cm$^{-2}$), respectively. **(b)** Part of **(a)** between DP and FFP peaks for electron doping is magnified and plotted as a function of $\phi_0/\phi$, where the $n$-independent oscillations near DP peak are calibrated by the uniform integer-valued $\phi_0/\phi$ (from 8 to 20, marked by the red dotted lines), while the left and right ends of the $n$-dependent are calibrated by different $\phi_0/\phi$ (left: from 15 to 23, marked by green dotted lines; right: from 14 to 30, marked by the orange dotted lines). **(c)** Color plot of $R_{xx}^{III}$ as a function of $n$ and $B$ ($V_{bg}$ is fixed at 20 V), where the yellow arrows mark the

$n$-dependent oscillations near FFP peak ($n \sim 1.05\times10^{13}$ cm$^{-2}$). **(d)** Part of **(c)** near FFP peak for electron doping is magnified and plotted as a function of $\phi_0/\phi$, where the left and right ends of the $n$-dependent oscillation are calibrated by different $\phi_0/\phi$ (left: from 15 to 31, marked by green dotted lines; right: from 14 to 30, marked by the orange dotted lines). **(e)** Cartoon of TBG sheet subject to external magnetic fields (indicated by flux quanta) where local angle disorder induced by line defects (left Inset) results in lateral heteromoiré superlattices with different wavelengths in region II, and lattice strain (right Inset) contributes to in-plane heteromoiré supercells with slightly different wavelengths or twist angles formed in region II and III.

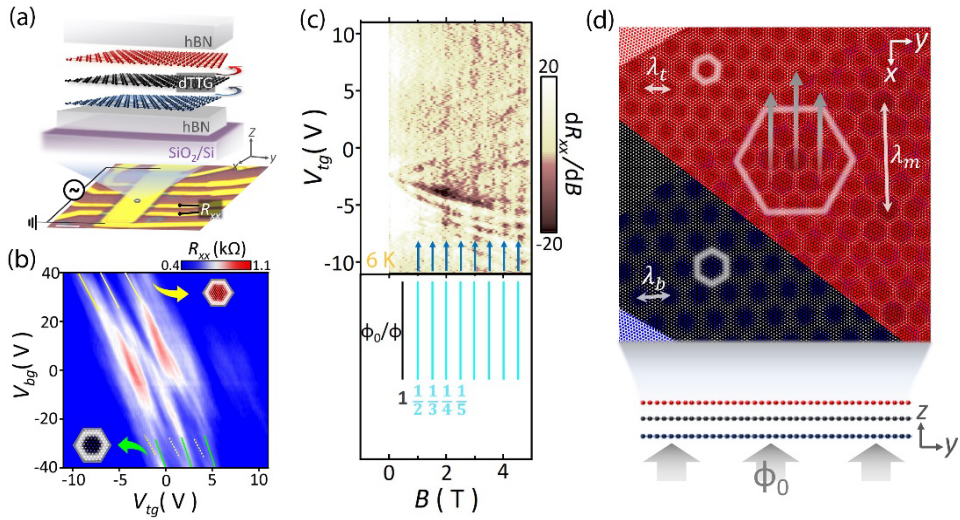

**Figure 4 | BZ oscillations governed by fractional MBSs in vertical heteromoiré superlattices with a mega-sized superlattices. (a)** Optical image and measurement configuration of the dual-gated dTTG superlattice Hall bar device T1, where schematic of the device geometry is illustrated. Scale bar, 2 μm. **(b)** Color plot of the longitudinal resistance $R_{xx}$ as a function of $V_{tg}$ and $V_{bg}$, respectively, where the yellow solid lines mark the DP and two SDP peaks (for hole and electron doping) assigned to the top TBG

moiré supercell (indicated by the yellow arrow and the red hexagon), while the green solid lines marks the peaks correspond to the bottom TBG superlattice (indicated by the green arrow and the black hexagon). Yellow dashed and solid lines have the same slope. **(c)** Upper: $dR_{xx}/dB$ as a function of $V_{tg}$ and $B$, exhibiting the oscillation in period $\Delta B \sim 0.5$ T ($V_{bg}$ is fixed at 40 V). Lower: Distribution of fractional MBSs in the corresponding superlattice with wavelength of $\lambda \sim 98.2$ nm. **(d)** Cartoon of the dTTG sheet subject to external magnetic fields (indicated by flux quanta) where the top and bottom TBG superlattices are stacked vertically and form the vertical heteromoiré superlattices, resulting in a new mega-sized superlattice dominating the oscillatory behavior (indicated by the flux quanta passing through it).